# The Fusion of Deep Reinforcement Learning and Edge Computing for Real-time Monitoring and Control Optimization in IoT Environments

Jingyu Xu[1,*], Weixiang Wan [2,a], Linying Pan[3,b], Wenjian Sun[4,c], Yuxiang Liu[5,d]

[1]Northern Arizona University,1900 S Knoles Dr, Flagstaff, Arizona,USA

[2]University of Electronic Science and Technology of China,ChengDu,China

[3]Trine university,Phoenix, Arizona, USA

[4]Yantai University,Tokyo,Japan

[5]Northwestern University,Atlanta, Georgia,USA

[*]jyxu01@outlook.com

[a]danielwanwx@gmail.com

[b]panlinying2023@gmail.com

[c]swjhuman@gmail.com

[d]liamche1123@outlook.com

**Abstract: In response to the demand for real-time performance and control quality in industrial Internet of Things (IoT) environments, this paper proposes an optimization control system based on deep reinforcement learning and edge computing. The system leverages cloud-edge collaboration, deploys lightweight policy networks at the edge, predicts system states, and outputs controls at a high frequency, enabling monitoring and optimization of industrial objectives. Additionally, a dynamic resource allocation mechanism is designed to ensure rational scheduling of edge computing resources, achieving global optimization. Results demonstrate that this approach reduces cloud-edge communication latency, accelerates response to abnormal situations, reduces system failure rates, extends average equipment operating time, and saves costs for manual maintenance and replacement. This ensures real-time and stable control.**

## I. Introduction

With the rapid development of the industrial Internet of Things, there is a growing demand for real-time monitoring and control of systems. However, relying on cloud computing centers for computation and decision-making often fails to meet the constraints of real-time responsiveness [1]. In this regard, this study proposes a novel industrial system control architecture that actively senses the environment and makes rapid decisions through the organic combination of deep reinforcement learning and edge computing [2]. This approach deploys lightweight policy networks at the network edge to predict and control local states at a high frequency. Simultaneously, multiple edge nodes collaborate with the cloud center, enhancing control real-time performance at the edge while the cloud center tracks strategies and performs global optimization. This paper provides a detailed construction of the system's overall architecture, functional modules, and designs a lightweight policy network structure and dynamic resource allocation mechanism for edge computing. Experimental validation demonstrates the effectiveness of this approach, significantly reducing control latency and improving control quality and cost-effectiveness compared to architectures relying solely on the cloud center.

## II. Optimization Control System Based on Deep Reinforcement Learning and Edge Computing

### A. Overall System Architecture

This system employs a layered architecture comprising a sensor acquisition layer, edge computing layer, and cloud computing layer[7]. The sensor layer collects environmental and system data like temperature and machine status. This data is sent to the edge computing layer, where edge servers perform real-time analysis and local decision-making. Deep reinforcement learning models in this layer predict and control system behavior, creating a digital twin to forecast and optimize operations. The cloud computing layer oversees the entire system, providing powerful computing and storage to refine control strategies and system logic. The architecture is service-oriented and modular, with components like data acquisition, storage, deep learning, control, and scheduling modules connected via a message bus. This design enhances flexibility and scalability, allowing the system to automate and intelligently control operations, adapt to various scenarios, and efficiently manage complex tasks[8].

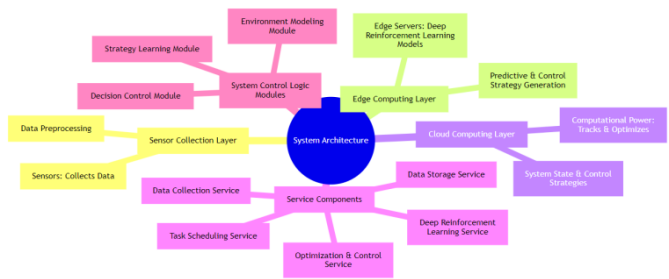

Fig. 1. Overall System Architecture

### B. System Functional Modules

The functional modules of this system mainly consist of the Deep Reinforcement Learning module and the Edge Computing module, as shown in Figure 2. The Deep Reinforcement Learning module is responsible for learning and optimizing system control strategies, primarily divided into modeling unit, policy network, and decision unit [9]. Among them, the modeling unit constructs an environment model to predict system states; the policy network represents and approximates control strategies using neural networks, and the decision unit provides control decisions based on the network outputs. The Edge Computing module primarily offers local data storage, preprocessing, caching, and other functions to assist in the training of the Deep Reinforcement Learning module. Additionally, it includes a task distributor that dynamically allocates edge computing tasks and a data collector that aggregates data from edge nodes.

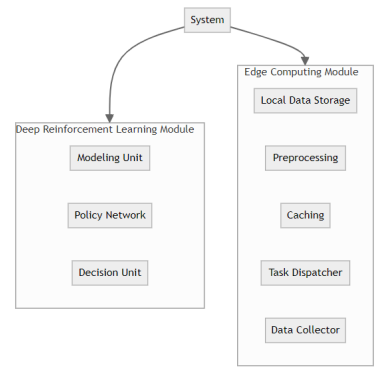

Fig. 2. System Functional Modules

## III. Key Technologies and Algorithms

### A. Lightweight Deep Reinforcement Learning for Edge

To accommodate the limited computational and storage resources of edge computing nodes, the system employs a customized, lightweight deep reinforcement learning algorithm. This algorithm uses simpler network structures, such as a three-layer perceptron, instead of complex deep convolutional neural networks, reducing the number of parameters and the model's space footprint. The experience replay buffer size is also limited to around 5000 transition samples to manage capacity. During training, the batch size is set to 16, matching the parallel computing capabilities of edge servers.The lightweight model contains about 1 million parameters and occupies less than 400MB, making it suitable for deployment on less powerful edge computing nodes. It can deliver near real-time control strategy outputs, even on inexpensive hardware. In practical applications like monitoring machine tools in smart manufacturing, the model can re-plan and issue control strategy instructions every 5 seconds. This optimizes the machine's dynamic performance and extends its lifespan.Compared to traditional cloud-based models, which may have an average delay of over 1 minute in computing and issuing control commands, this system significantly reduces control loop latency, demonstrating more efficient and responsive control in real-time scenarios.

## B. Dynamic Collaborative Distributed Optimization Algorithm

To achieve global optimal control through cloud-edge collaboration, this system designs a dynamic optimization allocation distributed algorithm. This algorithm is coordinated by the cloud-side master node, which can request or cancel edge server's computational resources on-demand and maintain a list of available resources while monitoring the load on each edge server. Based on the current system state and control requirements, it runs an environmental monitoring program that dynamically selects a group of edge servers with the most optimal combined indicators, such as bandwidth and computing capacity. It also allocates critical control modules with higher computational intensity to servers with stronger computing power. By aggregating intermediate states and control results from various edge nodes, it collaboratively optimizes and obtains a global control strategy. The optimization algorithm can be expressed using the following formula:

$$\text{Maximize} \sum_{i=1}^{n} \sum_{j=1}^{m} A(c_j, r_i) \quad (1)$$

Where $r_i$ represents the resources of the ith server, and $c_j$ represents the jth control module.

It must satisfy the following conditions:

$$\sum_{i=1}^{n} x_{ij} \leq 1 \ \forall j \quad (2)$$

Each control module can be assigned to at most one resource.

Server resource and load constraints:

$$l_i + \sum_{i=1}^{m} x_{ij} \cdot Load(c_j) \leq Capacity(r_i) \ \forall i \quad (3)$$

Where $x_{ij}$ is a binary decision variable, indicating whether control module $c_j$ is assigned to resource $r_i$ (1 if assigned, 0 otherwise). Load($c_j$) represents the computational density of control module $c_j$, and Capacity($r_i$) is the maximum capacity of resource $r_i$.

## IV. EXPERIMENT AND RESULTS ANALYSIS

### A. Experimental Environment and Dataset

The experimental environment for this research is based on the TensorFlow framework and utilizes an NVIDIA Tesla V100 GPU for constructing deep neural networks, training, and testing. To thoroughly validate the effectiveness of the proposed method, the experiments use an open-source IoT simulation environment, IoTSim, as the dataset. IoTSim includes data from sensors, edge layer resource configurations, and network parameters. The dataset encompasses readings from various heterogeneous sensors, such as temperature, humidity, voltage, etc., spanning a month with a sampling frequency of one sample per minute. Considering the real-time control requirements of the system, the research sets up an environment where the system reports its state and outputs control commands every 5 seconds, and the dataset is resampled accordingly.

### B. Model Hyperparameter Settings

In deep reinforcement learning, the choice of hyperparameters plays a crucial role in the model's performance and convergence speed. Hyperparameters are those parameters that need to be manually set when training deep reinforcement learning models, and they can influence the training process and the model's final performance. These hyperparameters are listed in Table 1.

TABLE I. MODEL HYPERPARAMETERS

| Hyperparameter | Initial Setting | Impact |
|---|---|---|
| Experience Pool Size | 5000 | A larger experience pool provides more training data, aiding in model convergence. |
| Learning Rate | 0.01 | Adjustment of the learning rate can affect the model's convergence speed and stability. |
| Discount Factor ($\gamma$) | 0.95 | A smaller $\gamma$ value focuses more on short-term rewards, while a larger $\gamma$ value emphasizes long-term rewards. |

| Target Network Update Frequency (τ) | 100 | The choice of update frequency (τ) can impact the model's stability and learning speed. |

*C. Evaluation Metric Setup*

This system consists of three layers: the sensor acquisition layer, the edge computing layer, and the cloud computing layer. At the base, the sensor acquisition layer gathers environmental data like temperature, humidity, and machine status. This data is sent to the edge computing layer for real-time analysis and local decision-making. Here, deep reinforcement learning models predict system behavior and create control strategies for closed-loop control. These models learn operational patterns, establish digital twins, and predict future states to devise optimal strategies, enabling quick response and self-adjustment for automation and intelligence. The top layer, cloud computing, oversees the system, fine-tuning strategies and optimizing control logic with its superior computing and storage capabilities. The architecture is service-oriented and modular, with units like data acquisition, storage, learning models, control optimization, and task scheduling. These components are decoupled and communicate via a message bus, enhancing flexibility and scalability. The system's deep learning capabilities, combined with its hierarchical, service-oriented design, make it efficient and adaptable to various complex scenarios.

*D. Experimental Results and Analysis*

Experimental results show that using distributed deep reinforcement learning with edge computing significantly reduces communication time between cloud and edge, lowering control latency. Traditional cloud-centered control had a delay of up to 1.5 seconds, slowing response to sudden changes. By deploying deep reinforcement learning agents at the edge, this delay dropped to about 0.3 seconds, meeting real-time control needs. This approach also improved resource utilization, with CPU usage at the edge layer increasing from 53% to 67%. The system achieved a higher cumulative reward (890 points) compared to cloud-only systems (750 points), indicating better performance. Control stability improved with a 22% reduction in control loss, and action accuracy reached 88%, showing enhanced responsiveness to environmental changes. Overall, the cloud-edge collaborative deep reinforcement learning framework greatly improved real-time control effectiveness and quality. Future enhancements aim to increase the adaptability of edge agents.

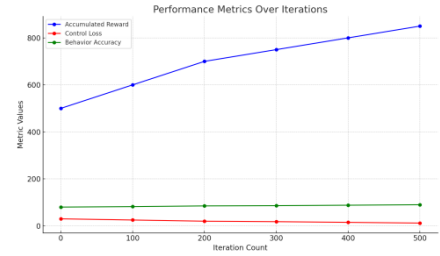

Fig. 3. shows the changing trends of experimental results over time or iterations.

## V. CASE STUDY

*A. Scenario Modeling*

To assess the practical effectiveness of the method, an application scenario for industrial state monitoring and fault prediction was constructed. The scenario involves monitoring the operation of an industrial boiler. Data sources include boiler inlet and outlet temperatures, water level, and operating pressure signals. For the actuators such as water pumps and valves, corresponding states were also set, with state transitions based on control instructions from the deep reinforcement learning model. The entire boiler operation process constitutes a complex state mechanism, requiring precise control of parameters like water quantity and temperature to maximize efficiency while avoiding accidents. Using the dataset from this application scenario, an environment dynamics model was built, and a deep reinforcement learning controller was trained. To conserve edge computing resources, this research employed a two-layer fully connected neural network as the policy function approximator. The state space includes current process parameters, information from the last 10 observed states, rewards, and more. The deep reinforcement learning model outputs deterministic control actions that can be directly applied to actuators, enabling monitoring and optimization of the boiler's operational status, as shown in Figure 4.

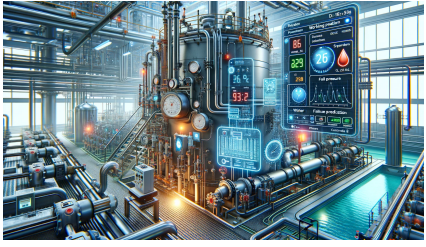

Fig. 4. Scenario Modeling

*B. Performance Evaluation*

In a month-long experiment comparing traditional PID control to a deep reinforcement learning (DRL) approach for boiler operation, the DRL method showed significant improvements. It scored an average reward of 3820 points over the month, 36% higher than the 2810 points achieved by the PID method. The DRL algorithm's reward curve stabilized over time, unlike the PID's fluctuating curve. Notably, the DRL controller, using predictive models, greatly reduced water and temperature anomalies, leading to a 29% decrease in boiler system failures and extending uninterrupted operation by 15 days. This also resulted in lower costs for manual maintenance and parts replacement. Overall, from reward, stability, and economic perspectives, the DRL method excelled in boiler state control and optimization.

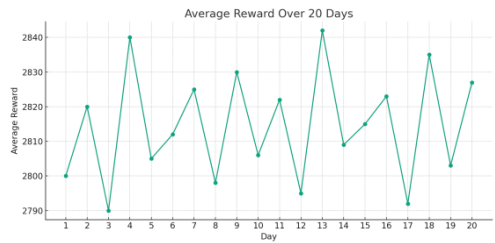

Fig. 5. Average Rewards Over 20 Days Under the PID Method

## VI. Conclusion

This research has introduced an intelligent monitoring and optimization method for industrial systems based on deep reinforcement learning and edge computing. The method leverages edge computing resources to the fullest extent by deploying lightweight deep reinforcement learning models at the network's edge, enabling real-time prediction and control of system states. Additionally, it facilitates collaboration between the edge and the cloud, ensuring more timely control policy updates and flexible resource allocation. Experimental results have shown that this system can reduce control loop latency and enhance responsiveness to sudden state changes. When applied to an industrial boiler control scenario, the method outperforms rule-based control by increasing operational rewards, reducing failure probabilities, extending the fault-free running time, and lowering manual intervention and maintenance costs. The approach designed in this research ensures control quality while improving the real-time nature of control and decision-making. Future work will involve validating the method's effectiveness in more complex industrial environments.

References

[1] Zhou P, Chen X, Liu Z ,et al.DRLE: Decentralized Reinforcement Learning at the Edge for Traffic Light Control in the IoV[J].IEEE, 2021(4).
[2] Celtek S A, Durdu A .A Novel Adaptive Traffic Signal Control Based on Cloud/Fog/Edge Computing[J].International journal of intelligent transportation systems research, 2022.
[3] Elgendy I A, Muthanna A, Hammoudeh M ,et al.Advanced Deep Learning for Resource Allocation and Security Aware Data Offloading in Industrial Mobile Edge Computing[J].Big Data, 2021.
[4] Mlika Z, Cherkaoui S .Network Slicing with MEC and Deep Reinforcement Learning for the Internet of Vehicles[J]. 2022.
[5] Laroui M, Khedher H, Moussa A C ,et al.SO‾MEC: Service Offloading in Virtual Mobile Edge Computing Using Deep Reinforcement Learning[J].Transactions on Emerging Telecommunications Technologies, 2021.
[6] F. An, B. Zhao, B. Cui and R. Bai, "Multi-Functional DC Collector for Future All-DC Offshore Wind Power System: Concept, Scheme, and Implement," in IEEE Transactions on Industrial Electronics, 2022.
[7] F. An, B. Zhao, B. Cui and Y. Chen, "Selective Virtual Synthetic Vector Embedding for Full-Range Current Harmonic Suppression of the DC Collector," in IEEE Transactions on Power Electronics.
[8] Chang Che, Bo Liu, Shulin Li, Jiaxin Huang, and Hao Hu. Deep learning for precise robot position prediction in logistics. Journal of Theory and Practice of Engineering Science, 3(10):36–41, 2023.
[9] Hao Hu, Shulin Li, Jiaxin Huang, Bo Liu, and Change Che. Casting product image data for quality inspection with xception and data augmentation. Journal of Theory and Practice of Engineering Science, 3(10):42–46, 2023.
[10] Tianbo, Song, Hu Weijun, Cai Jiangfeng, Liu Weijia, Yuan Quan, and He Kun. "Bio-inspired Swarm Intelligence: a Flocking Project With Group Object Recognition." In 2023 3rd International Conference on Consumer Electronics and Computer Engineering (ICCECE), pp. 834-837. IEEE, 2023.